# A Comprehensive Secure Protocol for all D2D Scenarios

Hoda Nematy

Malek-Ashtar University of Technology, Shabanlou, Babaee Hwy, Lavizan,Tehran.

## Abstract

*To fulfill two integral aims of abating cellular traffic and enhancing efficiency of cellular network, D2D is considered as a novel channel of communication. This form of communication has introduced for 4th cellular communication and enacts a significant role in the 5th generation. Four D2D communication scenarios defined in the references, includes direct D2D and relaying D2D communication both with and without cellular infrastructure. One of the major challenges addressing D2D protocols contributes to the fact that they have one single secure protocol that can adapt to the four scenarios. In the current study, we propose a secure D2D protocol based on ARIADNE. To authenticate and key agreement between Source and Destination, we employ LTE-A AKA protocol, further for broadcast authentication between relaying nodes TESLA was applied. In Contrary to the recent protocols, our proposed protocol has inconsiderable computation overhead and trivial communication overhead than SODE and preserve many security properties such as Authentication, Authorization, Confidentiality, Integrity, Secure Key Agreement, and Secure Routing Transmission. We check Authentication, Confidentiality, Reachability, and Secure Key Agreement of the proposed protocol with ProVerif verification tools.*

## 1. Introduction

Traditional cellular infrastructure does not allow cellular devices communicate with each other directly even in the close proximity. Such a strategy leads massive traffic to the cellular network and perchance, D2D communication has been introduced to overcome cellular traffic which enjoys high potential in providing more bandwidth and higher rates to the cellular network (1). Being more specific, D2D communication is considered as a peculiar approach in setting direct transmission between a Source and a Destination, though it provides scant interactions between cellular phones and the central nodes (i.e. eNodeB). It has been argued that D2D communication is used for close distances and cellular communication only for far enough distances (2). D2D first used in (3) for data transmissions between nodes and cellular communication. Some other researches (2–4) use D2D for cellular communication. Based on recent research, security has been remained as a continuing issue in the domain of D2D communication (5).Casting much light on this matter on hand, several security challenges including Authentication, Authorization, confidentiality, integrity… and a secure protocol have addressed all of them. The list of security solutions proposed by the recent references is presented in Table 1. It is apparent that non of the protocols preserves all security properties. The problem situation in (6) is based on a scenario that a user which covered by a healthy deactivated eNodeB intends to connect the cellular network and helps for communication and sharing secret keys. Having the connection processed in this protocol, two cryptic fields for each user have to be sent from eNodeBs to their eNodeBs neighbourhood; further, due to lack of information about which user may request communication and which user would respond to this request, multitude of communication overheads existed. Prior occurrence of the incident, every eNodeBs should send these fields to their users. Moreover, a user of healthy eNodeB may fall in the DOS attack by receiving too many requests from a malicious user. Sparkling much light, T. Ballan et al (7) use a Physically Unclonable Functions (PUF) to generate the secret key for each device. This circuit generates a unique value based on the idiosyncratic character of each D2D devices, afterwards uses this unique value with the public key of another device and Elliptic Curve Cryptography to generate a shared secret key which is used as an input value of the Salsa20 / 20 stream cryptographic function and creates a final message with the XOR operation of Salsa20/20 output and initial message as well. Besides the potential applicability, this method suffers few limitations, first, it is prone to man-in-the-middle attacks when the attacker is placed between the receiver and the transmitter and sends his public key to the parties. Second, it requires a PUF circuit in both devices. Looking at this matter on hand from another perspective, L. Wang et al (8) present a distributed group key sharing scenario based on computational Diffie-Hellman (CDH) key sharing protocol in the absence of cellular infrastructure. One line of shortcoming of this protocol is attributed to the fact that, it does not provide a security solution based on the presence of an attacker within the network. Each time a user adds or eliminates the group, a new session key should be created. A step higher. P. Gope protocol (9) verifies the identity of D2D devices inside the network coverage by a middle layer called the fog layer. This middle layer connects to the core network and can authenticate a device and share a secret key; furthermore, the device can also verify the received information by the fog layer without disclosing its identity information to this layer. It has been maintained that this method reduces the latency and enhances the mobility of end-users and also extends its utilization horizons even when a user is out of network coverage. A secure key exchange method between two D2D devices without network interference proposed in (10), that requires physical proximity of two devices prior communication. In the case of reusing a key, the security of communication will be severely compromised. It is also possible to reveal the key if one of the devices is infected with malware. In (11) a secure protocol for secure communication between eNodeB and GW was proposed. A summary of the security solutions of references shows in table 1. Our proposed protocol uses ARIADNE with TESLA (12) and LTE-A key distribution system and the coverage encompasses all four communication scenarios including D2D direct and indirect with and without cellular coverage (See Figure. 1). Concerning the mobile nature of D2D devices, as the encrypted message to the routing packet is added, a message would be transmitted opportunistically in the network. To be much concise, when users are mobile in D2D communication, they may change their location after each routing process and no longer participate in sending and receiving messages, therefore the routing procedure needs to be iterated. However, in our proposed protocol, redoing the routine operation is not required since through adding the encrypted message field to the routing package users have to participate in D2D as long as sending and receiving one packet process time.

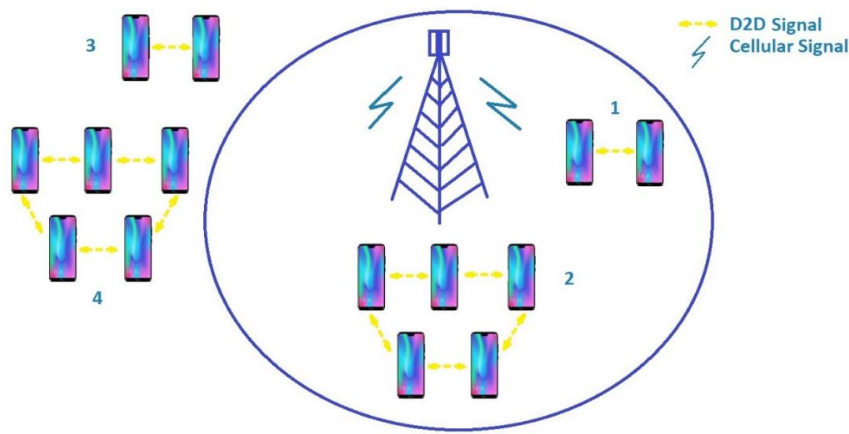

Figure 1. Four D2D Scenarios

As the cellular networks may become inaccessible in natural disasters, terrorist attacks, and other situations, in this scenario (out of coverage) the proposed protocol can override the network key agreement mode and use pre-shared keys. Applying the PUF circuit on D2D devices or the Diffie-Hellman key agreement protocol are the other two possible approaches for key agreement procedure. It should be pinpointed that such a key agreement is only restricted for the two factors of Source and destination, and relays do not require a pre-sharing secret key. Although using each pre-shared key method or a PUF circuit or key agreement protocols may reduce the security of our proposed protocol, we assume that in emergency situations, setting communication is of much great importance than abating communication security. Four D2D scenarios including, direct D2D with cellular infrastructure, direct D2D without cellular infrastructure, relaying D2D with cellular infrastructure and relaying D2D without cellular infrastructure are illustrated in Figure. 1. The rest of the paper devotes itself to peruse the subject in various sections. Section two will look through the four D2D secure protocols and schematic including direct D2D Secure Protocol (DD2D), Relaying D2D Secure Protocol (RD2D), direct D2D Secure Protocol without Cellular Infrastructure (DD2DW), and Relaying D2D Secure Protocol without Cellular Infrastructure (RD2DW). In section 3 the secure protocols will be analysed from three viewpoints of Computation overhead, communication overhead, and security properties. Ultimately, the security properties of secure protocols and Confidentiality, Reliability, one-way and two-way Authentication, and Secure Key Agreement in two phases will proof with the ProVerif formal verification tool will be discussed.

Table 1. Security solutions in D2D communication

|  | Authentication | Authorization | Confidentiality | Integrity | Secure routing transmission | Secure key agreement | Non-repudiation |
|---|---|---|---|---|---|---|---|
| SOD (6) | - | - | + | - | + | + | - |
| LAAP (9) | + | + | - | - | - | + | + |
| Sec-D2D (10) | + | - | + | + | - | + | - |
| SDR (7) | - | - | + | - | - | + | - |
| CRA (8) | - | - | + | + | + | + | - |

## 2. FOUR SECURE PROTOCOLS

We have four different protocol though they share a similar basis. Initially, a source incepts a D2D communication to the Destination. In scenarios 1 and 3, the Source and Destination are in each others' vicinity and could receive information directly. But, in scenarios 2 and 4, the Source and Destination are not in each others neighbourhood and need the cooperation of other devices to transmit and receive information. In scenarios 1 and 2, we use the cellular advantage to distribute keys between the Source and the Destination since, all the devices including Source and Destination are in the cellular coverage. Nevertheless, in line with dwindling cellular signalling traffic, the TESLA broadcast authentication protocol is employed for the intermediate nodes (i.e. relays). In scenarios 1 and 2, having fulfilled the establishment of a D2D communication to a specified destination the Source sends a D2D request message including Source and Destination identity in a secure cellular channel to the MME. The MME monitors the validity of the message and authenticates the Source and Destination, and also checks if the destination is in the proximity of the Source or not. If all the above situations would be satisfied, MME builds a D2D session key and sends it to the Source in a secure cellular channel and accordingly, the Source begins D2D communication towards Destination.

In the current protocol has its basis on ARIADNE, the packaging field includes S, D, id, and t for Source, Destination, the ID of the message and time respectively. Gaining a much comprehensive picture, a part of the encrypted message along with the nonce is annexed. Furthermore, in line with fulfilling the evaluation of MACs, a distinct key is exerted instead of two key, since in ARIADNE the Source and destination have each others key and one key is sufficed. We use the key chain TESLA protocol for intermediate users and assume that there is a system in the network where the initial values of the user's TESLA key chain are broadcasted to the entire network, and it can be implied that every cellular device can authenticate received TESLA key. In the presence and absence of cellular network coverage, two binary scenarios would be resulted. When users are in the coverage scope of the cellular network, this can be done by cellular network control messages. Whereas, in the absence of cellular network coverage, we assume that users use the previous initial values when the cellular network was available. The following section is devoted to provide an in depth illustration with regard to the four scenarios.

## 2.1. Direct D2D Secure Protocol(DD2D)

In the current protocol, two D2D devices are in each others' vicinity and Source initiates a D2D communication by requesting the core network to establish a D2D. The protocol procedure, parameters and fields are presented in Figure 2 and Table 2. The following nine phases thoroughly delineate the procedure.

1. First, in establishing a D2D communication to a specified receiver, the Source sends a D2D request message including Source and the Destination identity, in a Secure cellular channel to the MME.
2. In the second phase, MME enacts two major roles of checking the validity of the message and authenticates the Source and the Destination, and also monitoring the accuracy proximity of the Source to the destination. In meeting the aforementioned provisos, MME builds a D2D session key and sends it to the Source through a secure cellular channel.
3. The Source builds a K' key based on the key K received from the MME with a self-created nonce, afterwards the message with the key 'K' is encrypted. Furthermore, to fulfill the aim of integrity property, the Source builds a message authentication key (MAC) with the key K from the message field including the D2D request, Source id, Destination id, nonce, package id, and time.
4. In the fourth phase, the Source broadcasts all the existing data in the input of the MAC function with the result of it and h_0, to the receiver.
5. Succeeding, after receiving the message, to avoid repetition, the Destination first checks the package id, and then checks the value of time which should not be so far in the past. If all the situations agreed, a D2D request including Destination identity and Source identity is sent to the MME.
6. If there was an earlier D2D request to Destination from the specified transmitter, MME authorized the Destination and Source. Following the authorization, MME accepts the request and sends the key K to the Destination in a secure cellular channel in the step of 7.
8. Destination after getting the key K from MME, starts to decrypt the MAC function. If all the values are equal to the values in the request, the validity of the message will be accepted by the Destination and then starts to decrypt the package by evaluation of the key K' with the key K and nonce which was in the request fields.

9. Ultimately, after successful decryption and evaluation, the Destination builds a message and sends it to the transmitter. Putting the value of D2D reply, Destination identity and Source identity in a MAC leads to structure integrity.

Table 2: Parameter description

| Parameter | Description |
|---|---|
| K | secure session key between Source and Destination |
| MAC_K(M) | message authentication code of the message (M) with the key (K) |
| H() | Hash function |
| Enk_k() | symmetric Encryption with key k |
| Dek_K() | symmetric Decryption with key k |

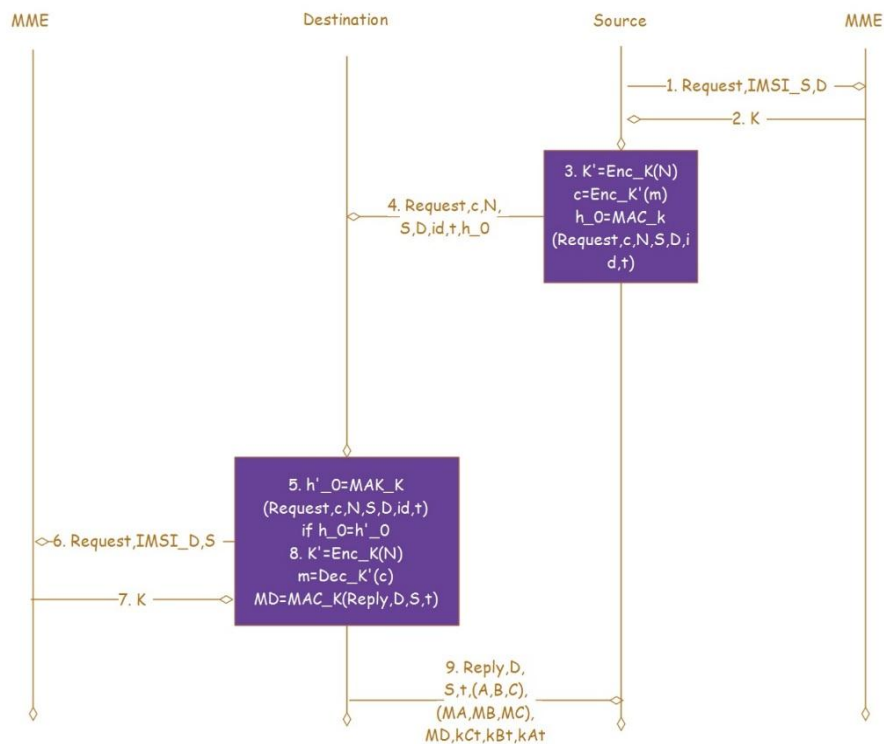

Figure 2. DD2D Protocol

## 2.2. Relaying D2D Secure Protocol(RD2D)

This protocol starts like DD2D by requesting a core network (i.e. MME). But in this scenario, the Source and the Destination are not in each others' vicinity and relaying nodes should participate to transfer information. The four phases of the protocol are the same as direct D2D protocol has been discussed in section 2.1. In line with gaining a thorough picture with regard to DD2D and protocol schematic, the following phases are presented (See Figure 3).

5. Device A after receiving the package from the transmitter checks the id for non-repudiation property and then checks the t value, this value should not be too far in the time. If all the values are true, it accepts the package and then evaluates the h_1 value based on h_0 and its identity. Then for integrity property calculates the MAC function on all the values on the previously received message plus its identity and h_1 with the key KAt which is the key to its TESLA key chain.
6. Then forwards all these values plus MA (MAC with the key KAt) towards Destination.
7. Device B also repeats all the processes above to the received message and makes MAC function based on KBt which is the key to its TESLA key chain.
8. Then forwards the package.
9. Device C repeats all the processes above and uses the key KCt for evaluation of MC. Then forwards the package in step 10.
Steps 11 and 12 are the same as steps 6 and 7 in DD2D which described in section 2.1.
13. Immediately after getting the key K from MME, Destination starts to evaluate the hash function chain including the MAC function with the key K has just been received. If the final chain is equal to the one in the request, the validity of the message is accepted by the receiver; and consequently the package decryption will be initiated and evaluated the key K' via the K key and nonce in the request fields.
14. After successful decryption and evaluation. The Destination builds a message reply and sends it to the transmitter. In achieving integrity properties, the destination puts the value of D2D reply, Destination identity, Source identity into a MAC function and encrypt it with the key K and sends it towards the transmitter.
15. In the penultimate phase, Device C receives a reply package and adds its TESLA key to the end of the package and moves it forwards.

In ultimate steps (16 and 17), B and A do the same proceeds as C and the package moves the transmitter forward.

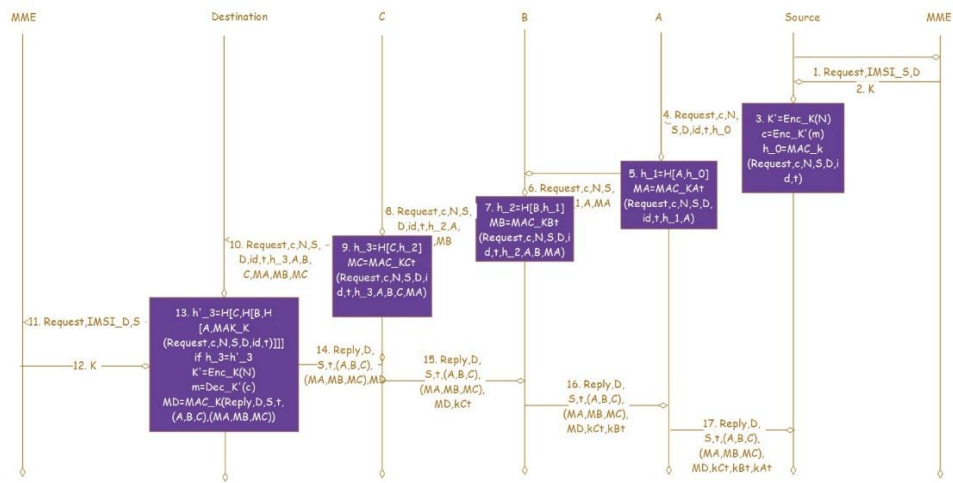

Figure 3: RD2D Protocol

## 2.3. Direct D2D Secure Protocol without Cellular Infrastructure (DD2DW)

Although this protocol is similar to the DD2D Protocol, steps 1,2, 6,7 do not exist because of lacking cellular infrastructure. To preserve confidentiality property, both Source and Destination have to use a key that set prior starting this form of communication. We suppose each device already exchanged secure key in a way such as key agreement procedures in (7,8). In the disaster situation, we suppose losing confidentiality is less important than losing vital communication at that time. So, we suppose each device has the potential to utilize its TESLA key if no other pre-distribution keys exist and could use no other procedures. The protocol description is illustrated in Figure 4 and the brass tacks explication is as follows.

1. Source starts to encrypt the message with the key K and then for integrity property puts the request D2D fields, encrypted file (c), nonce, source id, Destination id, package id, and t to the MAC function with its own TESLA key.
2. In the second phase, the Source sends the MAC function inputs with the result of MAC function (h_0) towards the Destination.
3. It should be pinpointed that prior receiving the TESLA key of the source, validation of the h_0 value is not feasible, though in the existence of the mutual key, it can decrypt the message [hence, in the case of emergency situation its better to decrypt the package and if the TESLA key which arrived from source failed to validate then the Destination withdraws the packet]. After package validation, Destination builds the reply package with the values of D2D reply, Destination id, source id, t, and the MAC value of these fields with the key KDt which is Destination TESLA key.
4. Hereafter, the MAC inputs with the MAC itself is sent towards the Source.

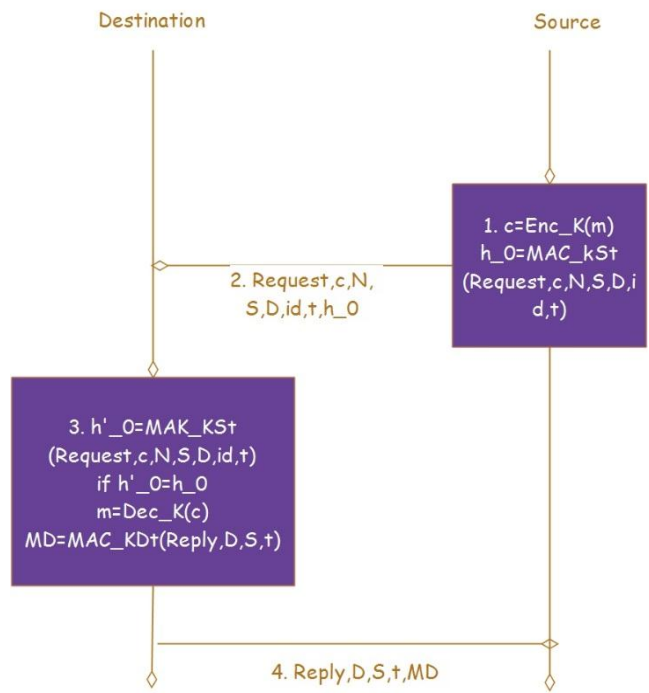

Figure 4: DD2Dw Protocol

## 2.4. Relaying D2D Secure Protocol without Cellular Infrastructure (RD2DW)

The structure of the current protocol is a combination of RD2D and DD2DW. As there should be no approximation between the Source and Destination, relaying nodes enact a vital role. It is pinpointed that since the cellular infrastructure is not available in this scenario, we presume that each device has already been exchanged secure keys. The rest of the protocol description is as follows (See Figure 5).

1. In the seedling stage, the Source begins D2D communication and encrypts the message with the key K. Then for integrity property puts the request D2D, encrypted file (c), nonce, Source id, Destination id, package id, and t to the MAC function with its own TESLA key.
2. In the next stage, the Source sends the MAC function inputs with the results of it (h_0) towards the Destination.

3. After receiving the package, device A checks the id for non-repudiation property and then checks the t value. Note that this value should not be too far in the past. If all the values were true, it accepts the package and then evaluates the h_1 based on h_0 and its identity. Then for integrity property, the Device calculates the MAC function on all the values on the previously received packet plus its identity and h_1 with the key KAt which is the key of its TESLA key chain.
4. In the fourth stage, all these values plus MA (MAC with the key KAt) are moved towards the Destination.
5. Device B also repeats all the above processes to the received packet and makes MAC function based on KBt, the key from its TESLA key chain, and accordingly creates MB.
6. Then it forwards the package received from A to the network, plus its id, B, and MB.
7. Device C repeats all the processes above and uses the key KCt for evaluation of MC.
8. Then forwards the package received from B to the network, plus its id, C, and MC.
9. In the existence of the mutual key, the Destination could validate the h_0 value before receiving the TESLA keys and decrypt the message as well. So, in the case of emergency situation its better to decrypt the package and if the TESLA keys arrived and the package failed to validate, then the Destination withdraws the package and informs all the network from the intruder. After validation of the package, the Destination builds the reply package with the values of D2D Reply, the Destination id, source id, and t, and the MAC value made of these fields with the key KDt which is Destination TESLA key.
10. The MAC inputs with the MAC itself is sent towards the source into the network.
11. After receiving a reply package, device C adds its TESLA key to the end of the package and moves it forward...

In steps 12 and 13, B and A do the same procedure as C and forward the package to the Source.

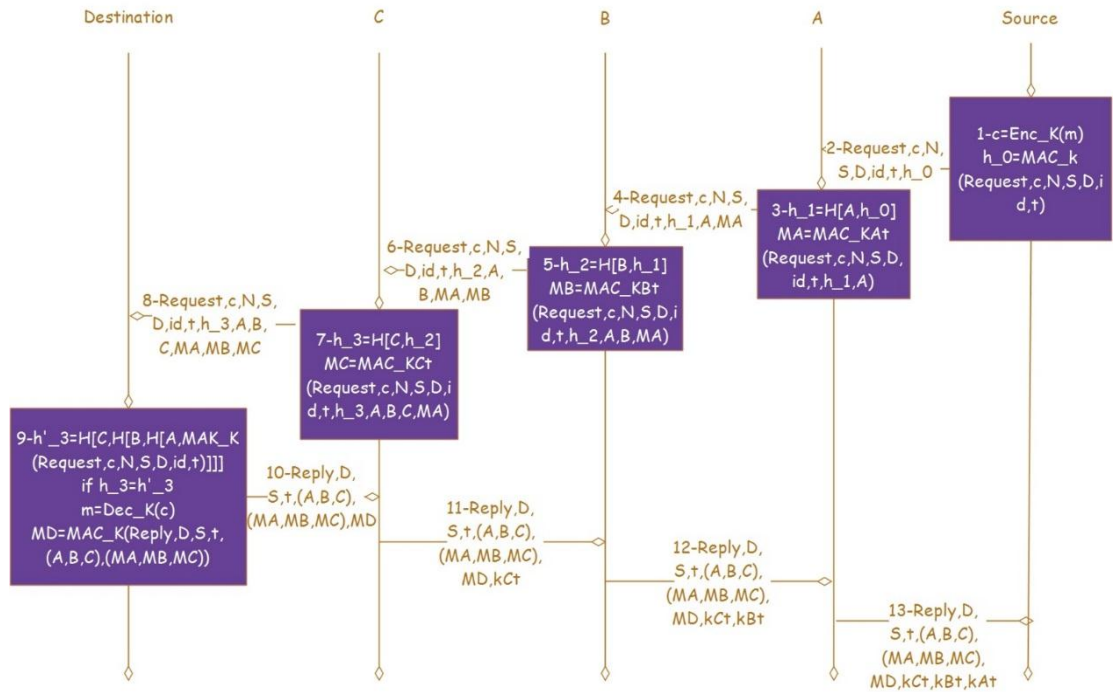

Figure 5: RD2DW protocol

## 3. ANALYSIS OF THE PROPOSED PROTOCOLS

As can be seen in Table 3, the role and the packet size of each node are considered as the determining factor in assessing the level of operations. In this Table, Enc and Dec are implemented for encryption and decryption; further, H is for hash value, Ks is for key size, and n is for the number of nodes including Source and Destination. We assume symmetric encryption with the output of 256 bits and also a hash function with the size of 256 bits, 4 bits for request, ti, I, and N and also 8 bits has its applicability for Source and Destination identities. Considering the number of nodes participating in D2D, the replay packet will have a different size. If we assume the maximum number of nodes is 20, the maximum packet size of Destination in the replay packet is 629 bytes and also the maximum packet size of intermediate nodes in request and replay packet respectively are 662 bytes and 629ks bytes.

Table 3: Operations and packet size in proposed protocols

| Device | operations | Packet size |
|---|---|---|
| The source in direct D2D | Enc+H | 544 bit |
| The source in relaying D2D | 2Enc+H | 544 bit |
| Destination in direct D2D | Dec+H | 286 bit |
| Destination in relaying D2D | Enc+Dec+nH | 28+(n-2)8+(n-1)256 bit |
| Intermediate node in the request packet | 2H | 28+(n-1)8+n256 bit |
| Intermediate node in the reply packet | - | 12+8n+(n-1)256+(n-2)Ks |

### 3.1. Computation overhead

In the proposed protocols, we use a symmetric function for encryption and decryption of the message and one for key and also a cryptographic hash function for each transmission. Therefore, there are two symmetric encryptions/decryptions and one cryptographic hash function evaluation for source and destination, and one cryptographic hash function evaluation for each relaying device. The computation cost of the proposed protocol is presented in Table 4. In line with a much more perception, Enc and Dec are for Encryption and Decryption, n is for the number of devices, H is a hash function, Mul is for multiplication, EO is for exponential operation, PA is for pairing, Div is for division and PO is for point multiplication.

Table 4: Computation cost of Secure protocols

| Secure protocol | Computation cost |
|---|---|
| SDGA (13) | $3(2n-1)PA + 5nEO + (4n-1)H + 2(2n-1)Mul$ |
| PPAKA (14) | $2(2n-1)EO + (n^2 + 3n - 4)H + (2n^2 - 3n + 1)Mul$ |
| GRAAD (15) | $2nPA + 7(3n-2)H + nEnc + nDec + 3(n-1)PO + 8(n-1)EO + 2(n-1)Mul$ |
| LRSA (16) | $6nPO + (13n - 7)H + (3n-1)Mul + 2Div$ |
| SeDS (17) | $2PA + (5n-2)EO + Dec + (2n+1)H + 4(n-1)PO + 2(n-1)Enc$ |
| DD2D | $3Enc + 3H + Dec$ |
| RD2D | $3Enc + (2n+1)H + Dec$ |
| DD2DW | $Enc + 3H + Dec$ |
| RD2DW | $Enc + (2n-1)H + Dec$ |

## 3.2. Communication overhead

In RD2D and RD2DW, the protocol has 2n packet transmission for each relay device (one for Request and one for Reply). Considering the feature, we can explicitly assert that the communication overhead of the proposed protocol is as equation 1.

$$Communication\ Overhead = \frac{T' \times M \times (2n+2)}{T} \qquad (1)$$

Sparkling much light in this matter on hand, $T'$ is the number of timeslots for the occurrence of D2D requests, the second core value is $M$, which is the number of D2D requests at each timeslot, and the number of applied devices is known as $n$. Since RD2D has the biggest communication overhead among three other proposed protocols, we compare the communication cost of RD2D with SODE (6). In SODE, two cryptic fields for each device has to be sent from each eNodeBs to each eNodeB neighbours. Also, two cryptic fields for each neighbours have to be sent to all the devices belongs to eNodeB. Another communication parts in SODE are from D2D request and D2D reply. These two communications are for key agreement between two devices in the network. Communication overhead of RD2D and SODE based on increasing the number of time slots when the number of eNodeBs are 2 and when are 7 in figure 6 and 7 respectively. It can be seen that the communication overhead increases as the number of nodes (n) increased. When the number of eNodeBs increased from 2 to 7, the communication overhead of SODE increases for about 3 times, but in RD2D the number of eNodeBs has no effect on the communication overhead. In another comparison, we check the change of the number of T' to communication overhead when M=1 and M=5 in figures 8 and 9 respectively. Concerning the quantification curve, as T' increases, the communication overhead increases and thus both protocols have more communication overhead if M increases to 5. It can be inferred that as the number of D2D requests increase the communication overhead increases as well. As illustrated in Figure 8 and 9, RD2D suffers less communication overhead than SODE. More to say, communication overhead in the slob of SODE is much more than RD2D.

Table 5. Parameters used in communication overhead simulation

| Parameter | value |
|---|---|
| n | 10 |
| T | 20 |
| T' | 10 |
| M | 1 & 5 |

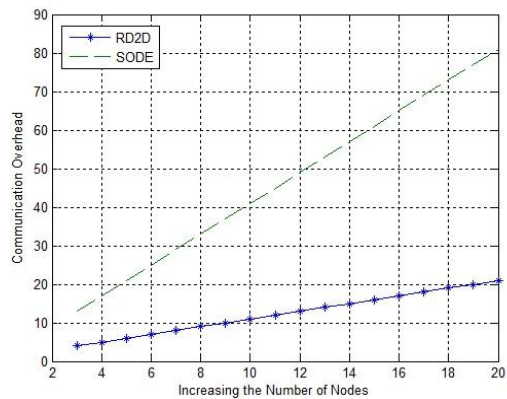

Figure 6. The Communication Overhead Vs the Number of Nodes when B=2

International Journal of Wireless & Mobile Networks (IJWMN), Vol.13, No.3/4, August 2021

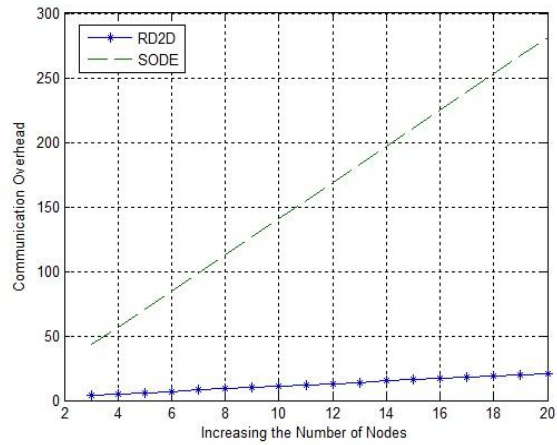

Figure 7. The Communication Overhead Vs the Number of Nodes when B=7

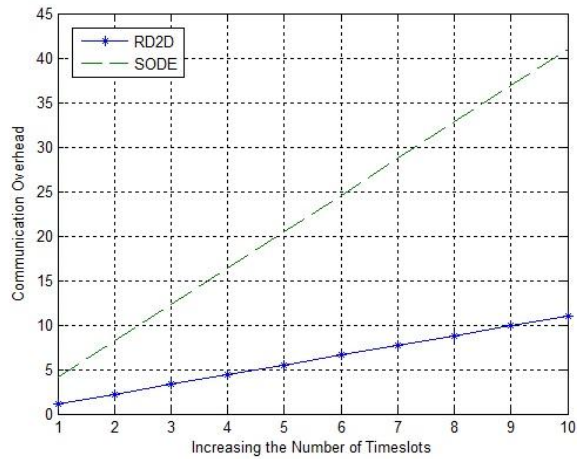

Figure 8. The Communication Overhead Vs the Number of Timeslots when M=1

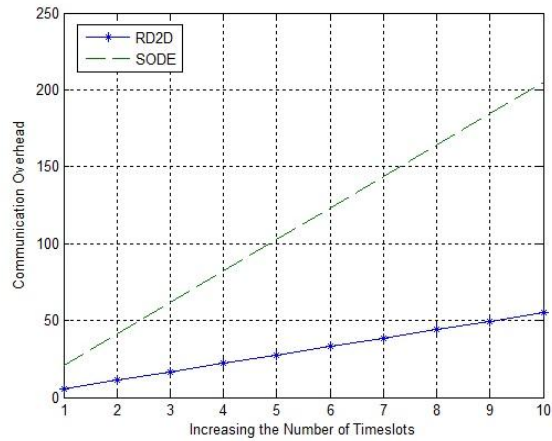

Figure 9. The Communication Overhead Vs the Number of Timeslots when M=5

### 3.3. Security properties of the protocol

In this section, the security properties of our protocols is demonstrate. Our proposed protocols consist of the integral components namely Authentication, Authorization, Confidentiality, Integrity, Non-repudiation, Secure routing transmission, Secure key agreement, and reachability. Making a bridge with other protocols, we will show two other security properties known as Secure key agreement and reachability in the ProVerif Section

1. Authentication and Authorization: This property is based on the cellular authentication and authorization process in cellular coverage scenarios (DD2D and RD2D). In the two other scenarios (DD2DW and RD2DW), authentication and authorization have their basis on the privacy of secret keys on each side. In meeting the two previous of decrypting the packet and evaluate the message, both sides (Source and Destination), are authorized. For this assumption, we suppose that no one reveals the key and the key saved in both devices securely.
2. Confidentiality: This property is gained by the encryption and decryption of the message based on the secret key, received from the MME. The MME is the trusted server which would not reveal the key K to anybody but authorized Source and Destination on each D2D communication. In DD2DW and RD2DW, the confidentiality of the message is based on the secrecy of the keys and key distribution system they used in the absence of cellular infrastructure.
3. Integrity: This feature is extracted by the hash values. If the destination evaluates the hash chain values and they are different from what was inside the packet, the integrity of the packet will be lost and the received packet should be waived. This property could be checked by the Source.
4. Non-repudiation: This property can be set by the packet id value in the request message which should be fresh. Also, the t value should not be too far in the past.
5. Secure routing transmission: Use of this property is restricted for RD2D and RD2DW, since these two protocols have routing part. As our proposed protocols are based on ARIADNE protocol, it prevents tampering with the attackers or comprised nodes and resists multitude Denial-of-Service attacks as well.

### 3.4. Proverif verification of RD2D protocol

It has been stated that ProVerif is considered as a formal tool to verify cryptographic protocols (18) and we check confidentiality, reachability and secure key agreement of RD2D by it. Input language of ProVerif supports channels with the "Dolev-Yao" ability attacker. This attacker model is very strong and has full control over the channel. Table 6 illustrates the security properties that we used for security validation of our proposed protocol.

Table 6. Security properties of the protocol used in ProVerif

| Security Property | | ProVerif |
|---|---|---|
| Confidentiality | | query attacker(m). |
| Reachability | | query event(mmeReachable()).<br>query event(hssReachable()).<br>query event(SourceReachable()).<br>query event(DestinationReachable()). |
| Authentication | One-way authentication | event acceptsServerClientA(bitstring,key).<br>event acceptsServerClientB(bitstring,key).<br>event acceptsServerClientC(bitstring,key).<br>event acceptsServerDestination(bitstring,key). |
| | One-to-one authentication* | event termDestination(bitstring,key). |
| Secure Key agreement | Running key | event SourceRunning(key).<br>event mmeRunning(key).<br>event DestinationRunning(key).<br>event ClientARunning(bitstring,key).<br>event ClientBRunning(bitstring,key).<br>event ClientCRunning(bitstring,key). |
| | Key agreement | event SourceCommit(key).<br>event mmeCommit(key).<br>event DestinationCommit(key). |

A unidirectional authentication is used to check authenticity (i.e. Source authenticates relaying devices and Destination). However, in one-to-one authentication, two sides of communication should authenticate each other (i.e. Source and Destination). It can be implied that, we use one-to-one authentication for Source and Destination and one-way authentication for relaying devices. We assume that MME is part of the core network and is trusted section and there is no need to be authenticated. We monitor the Secure key agreement procedure in two phases, running key and key agreement. In the phase of running key, a device uses a key and in the phase of key agreement, the other device agrees on the key used. As can be seen in Figure 10, ProVerif verifies all the security properties of RD2D.

```
ProVerif text output:

Starting query not event(SourceReachable)
goal reachable: end(SourceReachable)
RESULT not event(SourceReachable) is false.
Starting query not event(DestinationReachable)
goal reachable: end(DestinationReachable)
RESULT not event(DestinationReachable) is false.
-- Query event(SourceCommit(k_133)) ==> event(mmeRunning(k_133))
Completing...
200 rules inserted. The rule base contains 185 rules. 9 rules in the queue.
Starting query event(SourceCommit(k_133)) ==> event(mmeRunning(k_133))
goal reachable: begin(mmeRunning(kdf(n_128[imsi_127 = imsiS[!1 = @sid_35779],!1 = @sid_35780],k[!1 = @sid_35779]))) -> end(SourceCommit(kdf(n_128[imsi_127 = imsiS[!1 = @sid_35779],!1 = @sid_35780],k[!1 = @sid_35779]])))
RESULT event(SourceCommit(k_133)) ==> event(mmeRunning(k_133)) is true.
-- Query event(mmeCommit(k_134)) ==> event(SourceRunning(k_134))
Completing...
200 rules inserted. The rule base contains 185 rules. 9 rules in the queue.
Starting query event(mmeCommit(k_134)) ==> event(SourceRunning(k_134))
RESULT event(mmeCommit(k_134)) ==> event(SourceRunning(k_134)) is true.
-- Query inj-event(SourceCommit(k_135)) ==> inj-event(mmeRunning(k_135))
Completing...
```

Figure 10: ProVerif Verification of RD2D Protocol

## 4. CONCLUSION

In the context of the present study, we proposed four D2D secure protocols for four different scenarios (DD2D, RD2D, DD2DW, and RD2DW). To the best knowledge of the author, this is the first time a protocol has the capability to adapt four scenarios which are essential to D2D networks. The adopted framework was based on ARIADNE with TESLA. To fulfill the aim of Authentication and key agreement for the Source and Destination in RD2D and DD2D, we used LTE-A AKA protocol. Furthermore, TESLA and broadcast authentication protocol was implemented for key utilization in intermediate nodes. Enumerating the métiers, the protocol does not need pre-shared keys for these nodes. Relying on the results, in contrary to the recent literature, our proposed protocols have less computation overhead; moreover, RD2D suffers less communication perplexity in comparison to SODE protocol and it has more communication overhead among three other proposed protocols; thus, the others have less communication overhead than SODE as well .In a nutshell, we indicated that our protocol security features and proofs Confidentiality, Reachability, Authentication, Secure Key agreement with ProVerif formal verification tools. Casting much light on the fortes, our proposed protocols enjoys Authentication and Authorization, Confidentiality, Integrity, Non-repudiation, Secure routing transmission, Reachability, and Secure Key agreement with low communication and computation perplexity.

## AUTHORS

**Hoda Nematy** graduated with a master degree in Cryptography and Secure Communication from Malek Ashtar university of technology, Tehran, Iran. Currently, she is working as the R&D team manager in Pars Pooya Control Binalood Company.

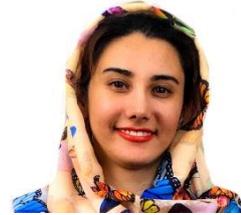